\documentclass[revsymb,bibnotes,aps,prl,amsmath,amssymb,reprint,superscriptaddress,floatfix]{revtex4-1}

\usepackage[]{graphicx}% Include figure files
\usepackage{bm}% bold math
\usepackage{float}
\usepackage{lineno}
\usepackage{hyperref}
\usepackage[all]{hypcap}
\usepackage{chngcntr}
\usepackage{siunitx}
\DeclareSIUnit{\dBm}{\deci\belmilliwatt}
\usepackage{autonum}

\usepackage{textgreek}
\usepackage{upgreek}
\usepackage{soul}
\usepackage{lipsum}
\usepackage{slashed}
\usepackage{amsmath}
\usepackage{amsfonts}
\usepackage{amssymb}%

\usepackage{balance}
\usepackage{color}
\makeatletter
\newcommand{\colorcaption}[2][]{%
  \begingroup%
  \renewcommand{\@caption@fignum@sep}{ (Color online). }%
  \caption[#1]{#2}%
  \endgroup%
}
\makeatother

\graphicspath{{Figs/}}
\begin{document}

\author{T. Hula}
\affiliation{Helmholtz-Zentrum Dresden-Rossendorf, Institute of Ion Beam Physics and Materials Research, Dresden, Germany}
\affiliation{Institut f\"{u}r Physik, Technische Universit\"{a}t Chemnitz, 09107 Chemnitz, Germany}
\author{K. Schultheiss}
\affiliation{Helmholtz-Zentrum Dresden-Rossendorf, Institute of Ion Beam Physics and Materials Research, Dresden, Germany}
\author{F.~J.~T. Gon\c{c}alves}
\affiliation{Helmholtz-Zentrum Dresden-Rossendorf, Institute of Ion Beam Physics and Materials Research, Dresden, Germany}
\author{L. K\"{o}rber}
\affiliation{Helmholtz-Zentrum Dresden-Rossendorf, Institute of Ion Beam Physics and Materials Research, Dresden, Germany}
\affiliation{Fakult\"{a}t Physik, Technische Universit\"{a}t Dresden, Dresden, Germany}
\author{M. Bejarano}
\affiliation{Helmholtz-Zentrum Dresden-Rossendorf, Institute of Ion Beam Physics and Materials Research, Dresden, Germany}
\author{M. Copus}
\affiliation{Center for Magnetism and Magnetic Nanostructures, University of Colorado, Colorado Springs, USA}
\author{L. Flacke}
\affiliation{Walther-Mei{\ss}ner-Institut, Bayerische Akademie der Wissenschaften, Garching, Germany}
\affiliation{Physik-Department, Technische Universit\"{a}t M\"{u}nchen, Munich, Germany}
\author{L. Liensberger}
\affiliation{Walther-Mei{\ss}ner-Institut, Bayerische Akademie der Wissenschaften, Garching, Germany}
\affiliation{Physik-Department, Technische Universit\"{a}t M\"{u}nchen, Munich, Germany}
\author{A. Buzdakov}
\affiliation{Helmholtz-Zentrum Dresden-Rossendorf, Institute of Ion Beam Physics and Materials Research, Dresden, Germany}
\author{A. K\'akay}
\affiliation{Helmholtz-Zentrum Dresden-Rossendorf, Institute of Ion Beam Physics and Materials Research, Dresden, Germany}
\author{M. Weiler}
\affiliation{Walther-Mei{\ss}ner-Institut, Bayerische Akademie der Wissenschaften, Garching, Germany}
\affiliation{Physik-Department, Technische Universit\"{a}t M\"{u}nchen, Munich, Germany}
\affiliation{Fachbereich Physik and Landesforschungszentrum OPTIMAS, Technische Universit\"{a}t Kaiserslautern, Kaiserslautern, Germany}
\author{R. Camley}
\affiliation{Center for Magnetism and Magnetic Nanostructures, University of Colorado, Colorado Springs, USA}
\author{J. Fassbender}
\affiliation{Helmholtz-Zentrum Dresden-Rossendorf, Institute of Ion Beam Physics and Materials Research, Dresden, Germany}
\affiliation{Fakult\"{a}t Physik, Technische Universit\"{a}t Dresden, Dresden, Germany}
\author{H. Schultheiss}\email{h.schultheiss@hzdr.de}
\affiliation{Helmholtz-Zentrum Dresden-Rossendorf, Institute of Ion Beam Physics and Materials Research, Dresden, Germany}

\title{Spin-wave frequency combs}

\begin{abstract}
We experimentally demonstrate the generation of spin-wave frequency combs based on the nonlinear interaction of propagating spin waves in a microstructured waveguide. By means of time- and space-resolved Brillouin light scattering spectroscopy, we show that the simultaneous excitation of spin waves with different frequencies leads to a cascade of four-magnon scattering events which ultimately results in well-defined frequency combs. Their spectral weight can be tuned by the choice of amplitude and frequency of the input signals. Furthermore, we introduce a model for stimulated four-magnon scattering which describes the formation of spin-wave frequency combs in the frequency and time domain.
\end{abstract}   

\maketitle

Frequency combs have been attracting interest in multiple fields of research in the last decades. For example, optical frequency combs have had great impact on metrology as they have led to a significant enhancement in the precision of frequency measurements \cite{Bellini2000, Jones2000, Fortier2019}. In solid state magnetism, several mechanisms are known to cause frequency spectra with equidistantly spaced spin-wave (SW) modes, by means of nonlinear interactions known as modulation instabilities \cite{Benner2000, Wu2004, Boyle1997, Kalinikos1991, Tsankov1994, ANSlavin1996, Staudinger1998}. These were often discussed in the time domain as they were found to cause soliton formation.
In addition, frequency mixing of two independently excited quasi-uniform modes has been studied in magnetic thin films leading to the generation of additional equidistantly spaced modes  \cite{Khivintsev2011, Marsh2012, 2Marsh2012}. Very recently, micromagnetic simulations have shown the generation of magnonic frequency combs by nonlinear three-magnon scattering of spin waves interacting with the intrinsic modes of magnetic skyrmions \cite{Wang2021}.

In microstructured magnetic waveguides, the process of four-magnon scattering (4MS) has been reported to populate a continuum of SW states in frequency ranges of up to several GHz around a strongly driven mode \cite{Suhl1957, Schultheiss2012, Hula2020}.
% In previous investigations, we shed light on the impact of nonlinear 4MS on the propagation characteristics of SWs leading to a significant increase of losses \cite{Hula2020}. 
4MS commonly describes the decay of two initial magnons, the quanta of SWs, into a pair of secondary magnons with different energies and momenta while conserving total energy and momentum \cite{Suhl1957, Krivosik2010}. 4MS is often considered parasitic and undesirable \cite{Scott2004, Demidov2009, Bauer2015} since it leads to additional damping of a strongly populated SW mode by coupling to the continuum of thermally populated states within the SW band \cite{Suhl1957, Schultheiss2012, Venugopal2020}. 
%Due to 4MS processes, a strongly populated SW mode loses energy by coupling to the continuum of thermally populated states within the SW band \cite{Suhl1957, Schultheiss2012}. 

In this letter, we take active control of 4MS in order to create tunable SW frequency combs. Those have significant impacts for both fundamental and applied physics as the superposition of frequencies, and associated wavevectors, allows one to create specific high-wavevector magnons which could not be readily created with typical geometries. In terms of applications, this frequency comb creates a series of microwave pulses in the time domain where the width of each pulse can be significantly adjusted by changing the amplitude of the initial signals.  Furthermore, our results experimentally demonstrate the efficient down-conversion of spin-wave frequencies which is important for the integration of magnonic devices into CMOS environments, an issue currently addressed by multiple complex approaches including modulation of optical frequency combs \cite{Moon2005}, superconducting qubits in cavities \cite{Fang2015}, and electronic circuits \cite{Yeh2014}. Finally, we explore the time-scale of the nonlinear responses, a topic not substantially addressed in earlier works. 

\begin{figure}
\centering
\includegraphics[width=8.6cm]{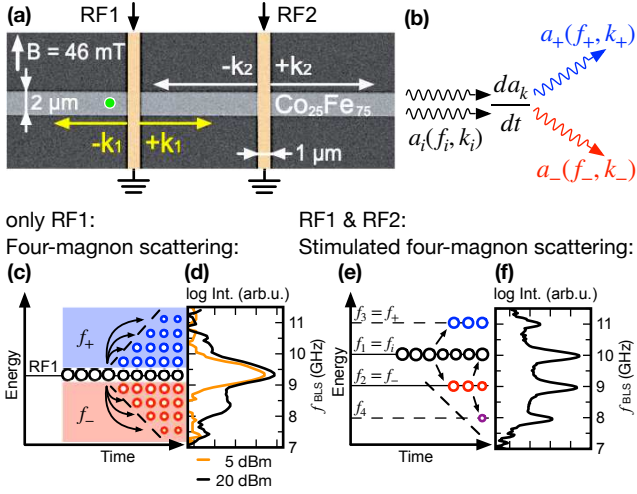}
\caption{(a) Scanning electron micrograph of the SW waveguide and the microwave antennas RF1 and RF2. The arrows $k_1$ and~$k_2$ illustrate the wave vectors of the SWs excited by RF1 and RF2, respectively. All measurements were performed in 1\,$\mu$m distance to the left edge of RF1 as indicated by the green dot.
(b) Schematic illustration of four-magnon scattering: two magnons with $(f_{i}, k_i)$ scatter into magnons with $(f_+, k_+)$ and $(f_-, k_-)$, at a rate defined by $da_k/dt$. (c) Energy-time diagram illustrating how spontaneous 4MS populates the SW band. (d) $\mu$BLS spectra measured in the SW waveguide only exciting with RF1 at two different powers. (e) Energy-time diagram for stimulated 4MS. (f) $\mu$BLS spectrum measured when exciting with RF1 and RF2 simultaneously.}
\label{fig:FIG1}
\end{figure}

We experimentally demonstrate the generation of such a frequency comb by exciting SWs in a Co$_{25}$Fe$_{75}$ micro\-conduit \cite{Schoen2016, Korner2017, Flacke2019} with a width of 2\,$\mu$m, a length of 60\,$\mu$m and a thickness of 30\,nm, as shown in Fig.~\ref{fig:FIG1}(a) \cite{Fabrication}. Two 1\,$\mu$m wide microwave antennas, referred to as RF1 and RF2, are patterned on top of the magnetic waveguide. Connecting the antennas to two microwave sources allows for the independent excitation of SWs at potentially different power levels $P_{1}$ and $P_{2}$. In the following, SWs generated by RF1 (RF2) will be referred to as mode $a_1$ ($a_2$) with frequency $f_1$ ($f_2$) and wave vector $k_1$ ($k_2$).
In all measurements, an external in-plane magnetic field of 46\,mT is applied which is not sufficient to fully saturate the waveguide perpendicular to the transport direction but still yields long SW decay lengths, as reported for a similar geometry in \cite{Flacke2019, Hula2020}. Note, however, that the non-collinearity at the edges of the sample is not essential for the emergence of 4MS, which is possible also in fully saturated specimens \cite{Lvov1994}.
%The dispersion relation of those SWs is quasi-linear in a wide wave-vector range which enables energy and momentum conservation for 4MS.  

Figure~\ref{fig:FIG1}(b) schematically shows the process of spontaneous 4MS \cite{Suhl1957, Schultheiss2012, Hula2020}. Using the notation of nonlinear SW theory \cite{Krivosik2010, Lvov1994}, we describe the current state (phase and oscillation magnitude) of a SW mode as a time-dependent complex amplitude variable $a_\nu (t)$ (with some mode index $\nu$), classical analoga to bosonic creation and annihilation operators $\hat{a}_\nu$ in second quantization. When a mode $a_i$ is excited above a certain threshold amplitude, its energy redistributes within the SW band via scattering of two initial magnons $a_i$ characterized by $f_i$ and $k_i$ into a pair of secondary magnons $a_+$ ($f_+$, $k_+$) and $a_-$ ($f_-$, $k_-$) under conservation of energy and momentum. 

The rates $\mathrm{d} a_k / \mathrm{d} t$ for these scattering processes, displayed in Fig.~\ref{fig:FIG1}(b), describe the change in amplitudes of the participating modes $a_k$ over time and can be derived using the Hamiltonian formalism for nonlinear SW dynamics \cite{Krivosik2010, Lvov1994}.
\begin{align}
\frac{\mathrm{d} a_k}{\mathrm{d} t} = &-i \omega(\mathbf{k}) a_k - \Gamma(\mathbf{k})(a_k - a_{k,\mathrm{th}}) \\
&+ i \gamma b_{\mathrm{RF}}e	^{-i \omega_{\mathrm{RF}} t} - i \frac{ \partial \cal{H}_{\mathrm{int}}}{\partial a^*_k} \label{eq:EQ1}
\end{align}

In Eq.~\ref{eq:EQ1}, the SW dispersion $\omega(\mathbf{k})$, the linear damping $\Gamma(\mathbf{k})=\alpha \omega(\mathbf{k})$ with $\alpha = 0.004$ typical for Co$_{25}$Fe$_{75}$, a phenomenological term for thermal population $a_{k, \mathrm{th}} = i \sqrt{N_{k, \mathrm{th}}}e^{-i \omega(k)t}$, the pump field $b_{\mathrm{RF}}$, and the interaction Hamiltonian $\cal{H}_{\mathrm{int}}$ are included. Since in our experiments $f_{i}/2$ is located in the SW band gap, resonant three-magnon scattering is prohibited and is therefore excluded \cite{Krivosik2010, Suhl1957}. Therefore, the lowest order of interaction to be considered for our calculations is the decay of the initial mode $a_i$ into the final states $a_+$ and $a_-$ and the corresponding Hamiltonian is written as:
\begin{equation}
{\cal{H}}_{\mathrm{int}} \simeq{\cal{H}}^{(4)} = \sum_{i + -} W_{ii,+-}a_{i} a_{i} a^\ast _+ a^\ast _-
\label{eq:EQ2}
\end{equation}
 
The rates given in Eq.~\eqref{eq:EQ1} depend on the coupling coefficients $W_{ii,+-}$ between the different modes as well as the amplitudes of the three modes $a_{i}$, $a_+$ and $a_-$ \cite{Krivosik2010}. As the coupling coefficients are inversely proportional to the wave-vector mismatch of the participating modes \cite{Krivosik2010, Hula2020}, the delay time for populating secondary modes via 4MS scales with the frequency spacing $\Delta f = |f_{\pm} - f_{1}|$, as illustrated by dashed lines in Fig.~\ref{fig:FIG1}(c). This delay time can be understood as the time needed to compensate for the intrinsic damping of the secondary states.

In order to study stimulated 4MS experimentally, we employ Brillouin light scattering microscopy ($\mu$BLS) \cite{Sebastian2015}. Figure~\ref{fig:FIG1}(d) shows the anti-Stokes side of $\mu$BLS spectra  that we measured when exciting SWs with antenna RF1 at a fixed frequency $f_1= 9.25$\,GHz but at two different pumping powers. At low powers (orange line), the SW spectrum shows a single symmetric peak at $f_1$. When increasing the pumping power above the threshold for 4MS (black line), a broadening of the peak is measured as a consequence of the energy redistribution caused by spontaneous 4MS.

If aiming at the generation of a SW frequency comb with a well-defined frequency spectrum, we need to actively increase the scattering rate for one specific scattering channel. Looking at Eqs.~\eqref{eq:EQ1},\eqref{eq:EQ2}, this can be achieved by populating one of the secondary modes (e.g. $a_{-}$) as a seed mode using antenna RF2 in addition to the excitation of the initial mode $a_{i}$ using antenna RF1 (or vice versa). As energy and momentum have to be conserved under 4MS, a third mode $a_{+}$ with $f_+ = f_{i} + \Delta f$ is generated simultaneously, as indicated in Fig.~\ref{fig:FIG1}(e). Thus, by actively populating the seed mode ($a_{-}$) we control all frequency channels of the scattering process. We refer to this process as stimulated 4MS, similar to the process of stimulated three-magnon scattering as shown in \cite{Schultheiss2019}.
%Since the generation of certain secondary modes $f_{2}$ and $f_3$ is actively controlled under these conditions, we refer to this process as stimulated 4MS. 
 %and thereby controlling all frequency channels of 4MS. The increasing amplitude of the secondary mode $a_{2}$ leads to an increase of the scattering rate (Eq.~\ref{eq:EQ1}, \ref{eq:EQ2}) from the initial mode $(a_1)$ into the already populated secondary mode $a_2$.
%By actively increasing the amplitude of a seed mode, the scattering rate, given by Eqs.~\ref{eq:EQ1},\ref{eq:EQ2}, from the initial mode $f_{1} (a_i)$ into the well defined mode $f_{2} (a_-)$ will increase. 

As depicted in Fig.~\ref{fig:FIG1}(e), higher order scattering processes then lead to the formation of a SW frequency comb. The secondary modes $a_{-}$ and $a_{+}$ themselves take the role of initial and seed modes in new scattering events. Therefore, higher order processes are stimulated, e.g., with two magnons $a_{2} = a_{-}$ scattering into modes $a_{i}$ and $a_{4}$ (with $f_4=f_- -\Delta f$). In Fig.~\ref{fig:FIG1}(e), this is illustrated by an additional mode (violet circle) which appears with equidistant frequency spacing but at a later time. The higher order processes need to compensate the intrinsic damping of the participating states, which causes an increasing delay time, as indicated by the slope of the dashed line.

%Again, this process needs to compensate the intrinsic damping of the participating states, which causes an increasing delay time for the onset of higher order modes resulting from the stimulated 4MS process. This delay time is indicated by the slope of the dashed lines in Fig.~\ref{fig:FIG1}(e). 

One example of a measured SW frequency comb is shown in Fig.~\ref{fig:FIG1}(f). The initial mode is pumped at $f_i=f_1=10$\,GHz using antenna RF1 while the seed mode is provided by antenna RF2 at $f_-=f_2=9$\,GHz. In addition to those two frequencies, the BLS spectrum shows very distinct peaks at $f_+ = f_3 =11$\,GHz and $f_4=8$\,GHz. 

\begin{figure}
\centering
\includegraphics[width=8.6cm]{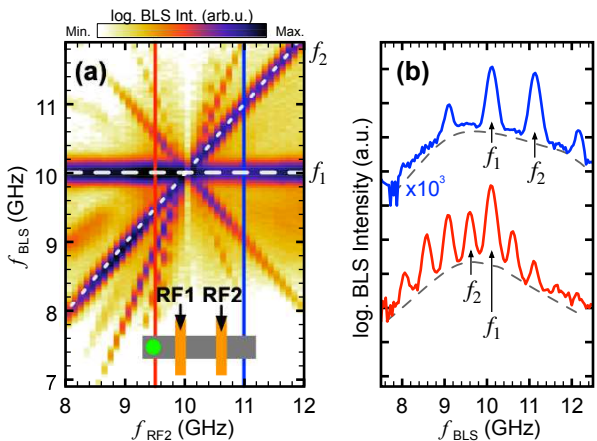}
\caption{(a) $\mu$BLS spectrum plotted as a function of the input frequency of RF2 while the frequency of RF1 was fixed at 10\,GHz. Both antennas were set to a constant input power of 10\,dBm. The frequency of RF2 was swept from 8 to 12\,GHz. (b) Examples of the BLS spectrum measured for RF2 set to 9.5\,GHz (red) and 11\,GHz (blue). The dashed lines indicate the continuum of states excited by 4MS.}
\label{fig:FIG2}
\end{figure}

In order to demonstrate the high degree of tunability of SW frequency combs, we measured $\mu$BLS spectra with a fixed frequency of the initial mode ($f_i=f_1=10$\,GHz) but varying frequency of the seed mode $f_2$ between 8\,GHz and 12\,GHz. The microwave powers of RF1 and RF2 were kept at constant values of $P_1=P_2=10$\,dBm. The resulting BLS spectra are plotted in Fig.~\ref{fig:FIG2}(a), with the SW intensity color coded on a logarithmic scale. The horizontal and diagonal white dashed lines indicate the applied frequencies $f_1$ and $f_2$. As $f_2$ is varied continuously to cross $f_1$, the frequency gap $\Delta f=| f_1-f_2|$ decreases up to $f_2=f_1=10$\,GHz and then increases again. Depending on the value of $\Delta f$, different numbers of indirectly excited SW modes are measured. Figure \ref{fig:FIG2}(a) demonstrates that the additional frequencies, observed in our experiments, strictly follow the spacing of the two driving frequencies $f_1$ and $f_2$ (dashed lines).

Figure~\ref{fig:FIG2}(b) shows individual BLS spectra measured for $f_2=9.5$\,GHz (red) and $f_2=11$\,GHz (blue), highlighting the different frequency spacing between neighbouring modes of 0.5 and 1\,GHz, respectively. In addition to the SW frequency combs, we measure a continuous background from the SW band between 8 and 12\,GHz, as depicted by dashed lines. This demonstrates that both mechanisms of spontaneous and stimulated 4MS take place at the same time. We note that, in addition to the coupling efficiency discussed earlier, the experimental observation of the continuum of SW states is limited towards lower frequencies by the SW band-gap (in our case at around 7\,GHz) and towards higher frequencies by the scattering geometry of the $\mu$BLS setup~\cite{Schultheiss2012, Sebastian2015, Hula2020}.

\begin{figure}
\centering
\includegraphics[width=8.6cm]{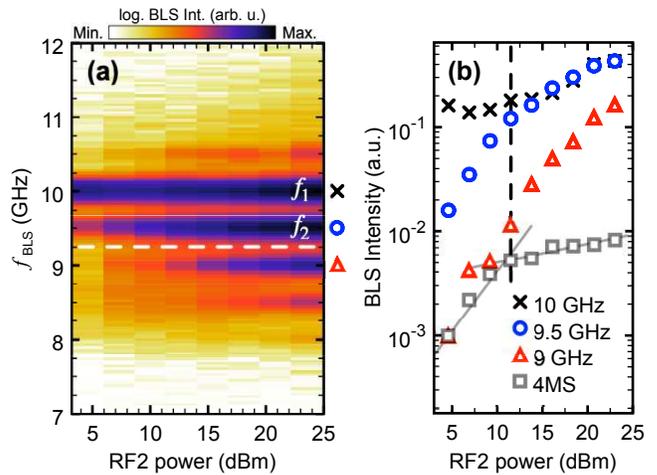}
\caption{(a) $\mu$BLS spectra measured as a function of the microwave power $P_2$ pumping   mode $a_2$ at a constant frequency $f_2=9.5$\,GHz. Mode $a_1$ is excited at a constant power $P_1=4.6$\,dBm and constant frequency $f_1=10$\,GHz. (b) Power dependent intensities evaluated for the different modes at 9\,GHz (triangles), 9.5\,GHz (circles), and 10\,GHz (crosses). The intensity of the continuum of spontaneous 4MS (squares) is extracted for 9.25\,GHz (dashed line in a). Lines are just a guide to the eye to highlight different slopes measured below and above 12\,dBm.}
\label{fig:FIG3}
\end{figure}

Previous studies have shown that the population of the magnon states excited via spontaneous 4MS increases with increasing pumping power \cite{Schultheiss2012, Hula2020}. In order to investigate the power dependence of stimulated 4MS, we fix the excitation frequencies of $a_1$ and $a_2$ at $f_1=10$\,GHz and $f_2=9.5$\,GHz, respectively. In addition, we fix the pumping power of mode $a_1$ at $P_1=4.6$\,dBm whereas mode $a_2$ is pumped with varying microwave powers $P_2$ ranging from 4.6  to 23\,dBm. The measured $\mu$BLS spectra as a function of $P_2$ are shown in Fig.~\ref{fig:FIG3}(a) with the BLS intensity color coded on a logarithmic scale. As expected, the intensity of $a_2$ continuously grows with increasing pumping power $P_2$ whereas only a minor increase of the intensity of $a_1$ can be observed. Besides the increase of intensity of $a_2$, additional modes at 8.5, 9, and 10.5\,GHz appear in the spectra, especially for larger values of $P_2$.
%With increasing pumping power of RF2 we observe a gradual increase of the intensity of the initial mode, an increase in the intensity of the continuum of states (broad background signal) and an increase in the intensity of the indirectly excited modes at 8, 8.5, 9 and \SI{10.5}{\giga\hertz}. 

The dependence of the SW spectra with respect to $P_2$ is discussed further with Fig.~\ref{fig:FIG3}(b), where the intensities extracted at 10 ($a_1$, crosses), 9.5 ($a_2$, circles), 9 (triangles), and 9.25\,GHz (4MS, squares) are plotted. The signal at 9.25\,GHz represents states in the continuum of the SW band that are only excited by spontaneous 4MS. 

Interestingly, the intensity of mode $a_1$ (crosses) is not constant despite its fixed excitation power of $P_1=4.6$\,dBm. Its intensity remains on the same level up to $P_2=11.5$\,dBm and then increases to values larger than those initially measured for $P_2=4.6$\,dBm. This development of the intensity of $a_1$ shows that we can distinguish different scattering regimes. 
%For $P_2<11.5$\,dBm, spontaneous 4MS dominates whereas for $P_2>10$\,dBm, stimulated 4MS with $a_2$ as the initial mode and $a_1$ as the seed mode becomes the pronounced scattering channel.

For $P_2<11.5$\,dBm, spontaneous 4MS of both modes $a_1$ and $a_2$ into the continuum of the SW band is the dominant scattering channel. Since $a_2$ is provided with more and more energy by the increasing microwave power, its intensity rises rapidly which directly leads to a strong increase of the intensity of the continuum of the SW band [squares in Fig.~\ref{fig:FIG3}(b)]. 
%However, mode $a_1$ is pumped at a constant power and increasing spontaneous 4MS leads to measurable losses.  

For $P_2>11.5$\,dBm, stimulated 4MS is the pronounced scattering channel and spontaneous 4MS plays a minor role only. Since mode $a_2$ [circles in Fig.~\ref{fig:FIG3}(b)] is pumped much stronger than $a_1$ (crosses), $a_2$ can be assigned as the initial mode and $a_1$ the seed mode. As a result, two magnons at $f_i=f_2=9.5$\,GHz scatter in one magnon at $f_+=f_1=10$\,GHz and one at $f_-=9$\,GHz (triangles). This not only leads to a rapid increase of the intensity at 9\,GHz above the level of 4MS but also to an increase of the intensity of $a_1$. Even though the intensity of $a_2$ continues to increase due to the larger pumping powers, it never exceeds the intensity of mode $a_1$. This is quite remarkable since at $P_2=23$\,dBm, $a_2$ is pumped significantly stronger than $a_1$ ($P_1=4.6$\,dBm) and still their intensities remain on the same level. This demonstrates the high efficiency of stimulated 4MS: most of the energy of mode $a_2$ is redistributed within the modes of the SW frequency comb. Hence, spontaneous 4MS is less pronounced as for $P_2<11.5$\,dBm and the intensity of the continuum [squares in Fig.~\ref{fig:FIG3}(b)] increases at a much lower rate. 

The mechanism of stimulated 4MS, outlined in Fig.~\ref{fig:FIG1}(e), presumes that the SW frequency comb is generated via a temporal cascade of 4MS events that populate discrete modes in the SW band. To investigate this temporal evolution experimentally, we use time-resolved BLS microscopy (TR $\mu$BLS) \cite{Sebastian2015}.

\begin{figure}
\centering
\includegraphics[width=8.6cm]{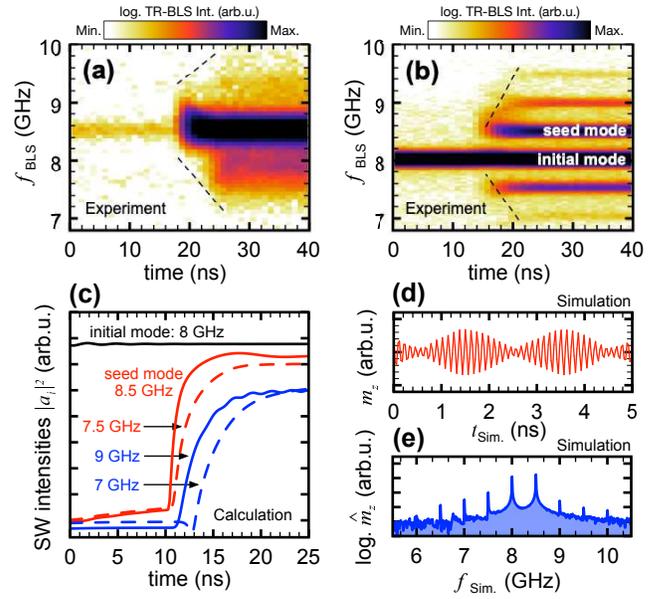}
\caption{(a) TR $\mu$BLS spectrum measured when exciting SWs with a single RF pulse at $f_1=8.5$\,GHz %(20\,dBm) 
(only the first 40\,ns are shown). (b) TR $\mu$BLS spectrum measured when continuously pumping the initial mode at $f_1=8$\,GHz 
%(11\,dBm) 
and exciting the seed mode at $f_2=8.5$\,GHz 
%(25\,dBm) 
by a microwave pulse. (c) Temporal profiles of modes of the frequency comb obtained by the analytic model in Eq.~\eqref{eq:EQ1}. (d) First 5\,ns of the 150\,ns-long time profile of $m_z$ and (e) FFT thereof, showing the simulated SW frequency comb.}
\label{fig:FIG4}
\end{figure}

First, we investigated the temporal response of the system to a single microwave pulse at $f_1=8.5$\,GHz. The pulse width was set to 150\,ns with a rise time of 3\,ns and a repetition period of 200\,ns. The first 40\,ns of the measurement are shown in Fig.~\ref{fig:FIG4}(a). The microwave pulse sets in after 17\,ns, accompanied by the characteristic frequency broadening of 4MS. Interestingly, it takes up to 6\,ns after the onset of the SW pulse for the frequency distribution to reach a dynamic equilibrium and fill states of the continuum with frequencies ranging from 7 to almost 10\,GHz. The diagonal dashed lines highlight that modes in the continuum with smaller frequency spacing to the directly excited mode are populated much faster than modes with larger frequency spacing [also see Fig.~\ref{fig:FIG1}(c)]. 
%The larger this frequency spacing, the longer the time delay of 4MS as already discussed with Figs.~\ref{fig:FIG1}(b)-(d).

Figure~\ref{fig:FIG4}(b) shows a TR $\mu$BLS spectrum measured for a continuously excited initial mode at $f_1=f_i=8$\,GHz and a pulsed seed mode at $f_2=f_+=8.5$\,GHz starting at 15\,ns with respect to the displayed time window. Here, the additional secondary mode $a_-$ at 7.5\,GHz is measured simultaneously to the seed mode. Higher order modes at 7, 9, and 9.5\,GHz are measured with larger time delays, depending on their frequency spacing to the initial mode. 

These measurements demonstrate that in both cases a certain time is needed to populate additional states via nonlinear interactions. The shorter delay times for the case of stimulated 4MS in Fig.~\ref{fig:FIG4}(b) suggest that the efficiency of energy transfer is enhanced when only discrete modes are excited using two RF sources.

Figure~\ref{fig:FIG4}(c) shows the results of the calculations based on Eq.~\eqref{eq:EQ1} for a continuously driven initial mode at 8\,GHz (black line) and a pulsed secondary mode at 8.5\,GHz setting in after 10\,ns. A set of seven rate equations was solved, referring to SW modes ranging from 6.5 to 9.5\,GHz in 0.5\,GHz steps. For simplicity, only five of these modes are shown by the modulus square $|a_i|^2$, which corresponds to the SW intensity and is proportional to the photon counts in our time-resolved BLS experiments. 
%In direct comparison, Fig.~\ref{fig:FIG4}(d) plots the temporal evolution of the same modes that 
We note that the delay times obtained analytically are in qualitative agreement with those obtained experimentally in Fig.~\ref{fig:FIG4}(b).

%When describing the mechanism for stimulated magnon scattering we implied that the coherent magnons injected using RF2 would master the coherence within the SW band, leading to selective amplification of indirectly excited magnon modes. Indeed, we confirmed this observation in the frequency domain, when showing the multiple SW peaks distributed across the SW band. This was further demonstrated by the much larger increase in the intensity of the indirectly excited SW modes compared to that of the continuous background excited via 4MS. 
In the time domain, a superposition of several modes translates into a time modulated amplitude which enables e.g. optical frequency combs to bridge the gap between optical and microwave frequencies \cite{Bellini2000,Jones2000}. In order to demonstrate a similar behaviour for the observed magnon frequency combs, we investigated the temporal evolution of the magnetization in micromagnetic simulations using MuMax3 \cite{Khivintsev2011, Arne2014} with material parameters and grid size listed in \cite{Simulation}.

In the simulations, the excitation field was strictly localized to the area of the two antennas in order to investigate the interaction of propagating waves and avoid any far-field excitations of the antennas. The excitation frequencies of RF1 and RF2 were set to 8 and 8.5\,GHz, respectively, with field amplitudes of 1\,mT.
%, sufficient to obtain a frequency comb. 
The temporal representation of the simulated SW frequency comb is displayed in Fig.~\ref{fig:FIG4}(d) in form of the amplitude of the dynamic component of the magnetization, $m_z(t)$. Here, we confirm a temporal modulation of the amplitude by 80\% with a 2\,ns period, which corresponds to the inverse of the 0.5\,GHz frequency spacing between the two input frequencies. Figure~\ref{fig:FIG4}(e) shows the fast Fourier transform (FFT) of $m_z(t)$. The resulting spectrum exhibits multiple SW modes in addition to the directly excited modes at 8 and 8.5\,GHz, resembling the measured SW frequency comb.

In conclusion, we observed the formation of spin-wave frequency combs generated by the nonlinear interaction of propagating spin waves excited at two different frequencies in a magnon conduit. 
A variety of spin-wave modes can be observed, whose frequency spacing and amplitudes were manipulated exemplarily by varying the frequency and power of one of the applied microwave currents.
The underlying mechanism can be understood in terms of stimulated four-magnon scattering, which was found to be the dominant process in the performed experiments. We propose that this effect can be utilized in any magnetic material and/or geometry as long as four-magnon scattering occurs.
\section{Acknowledgement}
This work was supported by the Deutsche Forschungsgemeinschaft within program SCHU 2922/1-1. MW, LL and LF acknowledge funding by the German research foundation (DFG) via projects WE5386/4-1 and WE5386/5-1.
\bibliography{SW_comb.bib}

\begin{thebibliography}{34}
\bibitem [{Fab()}]{Fabrication}%
  \BibitemOpen
  \href@noop {} {}\bibinfo {note} {The stripe of
  Co\textsubscript{25}Fe\textsubscript{75} was processed on
  Si/SiO\textsubscript{2} using electron beam lithography, magnetron sputtering
  and subsequent lift-off. The metallic thin film was deposited with a
  thickness of 30~nm. Two shorted-CPW gold antennas, processed via e-beam
  lithography, thermal evaporation and lift-off, were used as the microwave
  inputs RF1 and RF2.}\BibitemShut {Stop}%
\bibitem [{Sim()}]{Simulation}%
  \BibitemOpen
  \href@noop {} {}\bibinfo {note} {We modelled the conduit using 4096 cells
  along the long axis, 64 cells along the short axis and 1 cell along the film
  normal. For excitation we defined two rectangular regions across the stripe
  where dynamic out-of-plane fields could be applied. The position of these
  areas was similar to the fabricated sample. The following material parameters
  were used: $A_{ex}=26 \cdot 10^{-12}$~J/m\textsuperscript{3},
  $M_{sat}=1700\cdot10^3$~A/m and $\alpha=0.01$.}\BibitemShut {Stop}%
\end{thebibliography}

\end{document}